\documentclass[a4paper,11pt]{article}
\usepackage{pos}

\title{Hydrodynamic analyses of nuclear collisions in Landau and Eckart frames}

\author*[a]{Akihiko Monnai}

\affiliation[a]{
  Department of Mathematical and Physical Sciences, Japan Women's University\\ 
Bunkyo-ku, Tokyo 112-8681, Japan}

\emailAdd{monnaia@fc.jwu.ac.jp}

\abstract{
Relativistic dissipative hydrodynamic model at finite density is a promising tool for analyzing the dense QCD matter created in the beam energy scan experiments. The hydrodynamic frame can be chosen in the direction of energy flow, which is called the Landau frame, or conserved charge flow, which is called the Eckart frame. In this study, I investigate the stability and causality of full second-order relativistic hydrodynamic equations in the two frames. Then the effects of frame choice on hydrodynamic variables and experimental observables are estimated numerically and are found to be visible on the flow but limited on the rapidity distributions.
}

\FullConference{
  *** Particles and Nuclei International Conference - PANIC2021 ***\\
  *** 5 - 10 September, 2021 ***\\
  *** Online ***
}

\begin{document}
\maketitle

\section{Introduction}

The quark-gluon plasma (QGP) has been established as a relativistic fluid through experimental and theoretical efforts and discoveries in the past two decades. The dynamical QCD system created in high-energy nuclear collisions is quantified as a nearly-perfect fluid with extremely small yet finite viscosity.
The beam energy scan programs at BNL Relativistic Heavy Ion Collider have been performed to explore the QCD phase diagram in recent years. For a comprehensive analysis of the quark matter in such collisions, one has to take account of finite net baryon density and off-equilibrium corrections in the hydrodynamic model \cite{Monnai:2012jc,Denicol:2018wdp}. 

On the other hand, relativistic hydrodynamics is known to have the long-standing issue of frame fixing in a system with non-vanishing conserved currents. Two most common frame choices are the Landau frame, where the local rest frame of energy flow is used for the hydrodynamic flow \cite{Landau:1959}, and the Eckart frame, where that of conserved charge flow is used \cite{Eckart:1940te}. 

In this study, I investigate relativistic dissipative hydrodynamics in the Landau and Eckart frames \cite{Monnai:2019jkc}. The stability and causality conditions are examined 
in both frames. Numerical simulations are then performed to quantify the effects of frame choice on hydrodynamic variables as well as rapidity distributions in nuclear collisions.

\section{Stability and causality of relativistic dissipative hydrodynamics}

The relativistic fluid with a single conserved charge is considered. The hydrodynamic equations of motion consist of conservation laws $\partial_\mu T^{\mu \nu} = 0$ and $\partial_\mu N^\mu = 0$, and constitutive relations. Here the energy momentum tensor and the conserved charge current can be decomposed as
$T^{\mu \nu} = e_L u_L^\mu u_L^\nu - (P_L + \Pi_L) \Delta_L^{\mu \nu} + \pi_L^{\mu \nu}$ and 
$N^\mu = n_L u_L^\mu + V_L^\mu$ 
using the flow $u_L^\mu$ in the Landau frame.  The projection operator is defined 
as $\Delta^{\mu \nu} = g^{\mu \nu} - u^\mu u^\nu$. $e$ is the energy density, $P$ is the hydrostatic pressure, $\Pi$ is the bulk pressure, $\pi^{\mu \nu}$ is the shear stress tensor, $n$ is the conserved charge density, and $V$ is the diffusion current. The subscript $L$ denotes the quantities defined in the Landau frame. 
Hereafter the shear and bulk viscosities are neglected to focus on the vector dissipative currents.

The full second-order constitutive relations from the extended Israel-Stewart theory \cite{Israel:1979wp, Monnai:2010qp} reads
\begin{align}
V_L^\mu &= \kappa_V \nabla_L^\mu \frac{\mu}{T} - \tau_V (\Delta_{L})^{\mu}_{\ \nu} D_L V_L^\nu + \chi_V^a V_L^\mu D_L \frac{\mu}{T} \nonumber \\
&+ \chi_V^b V_L^\mu D_L \frac{1}{T} + \chi_V^c V_L^\mu \nabla^L_\nu u_L^\nu + \chi_V^d V_L^\nu \nabla^L_\nu u_L^\mu + \chi_V^e V_L^\nu \nabla_L^\mu u^L_\mu, 
\label{eq:diffusion}
\end{align}
in the Landau frame where $D = u^\mu \partial_\mu$ and $\nabla^\mu = \Delta^{\mu \nu} \partial_\nu$. $\kappa_V$ is the conductivity, $\tau_V$ is the relaxation time of the diffusion current, and $\chi_V$ are the other second-order transport coefficients.

Hydrodynamic modes can be obtained by taking the plane wave perturbation $\delta Q = \delta \bar{Q} e^{i(\omega t - kx)}$ from global equilibrium where $Q$ is a macroscopic variable \cite{Hiscock:1987zz}. The causality condition $| \partial \mathrm{Re} (\omega)/\partial k | \leq~1$ and the stability condition $\mathrm{Im} (\omega) \geq 0$ are satisfied when $\kappa_V\geq 0$ and $\tau_V \geq 0$ in the long wavelength limit. The detailed calculation can be found in Ref.~\cite{Monnai:2019jkc}.

The energy-momentum tensor and conserved charge current in the Eckart frame, on the other hand, are decomposed as 
$T^{\mu \nu} = e_E u_E^\mu u_E^\nu - (P_E + \Pi_E) \Delta_E^{\mu \nu} + W_E^\mu u_E^\nu + W_E^\nu u_E^\mu + \pi_E^{\mu \nu}$ and 
$N^\mu = n_E u_E^\mu$. Here $W^\mu$ is the energy dissipation current. 
The subscript $E$ is used to denote the quantities defined in the Eckart frame. The second-order constitutive relation is
\begin{align}
W_E^\mu &= - \kappa_W \bigg( \nabla_E^\mu \frac{1}{T} + \frac{1}{T} D_E u_E^\mu \bigg) - \tau_W (\Delta_{E})^{\mu}_{\ \nu} D_E W_E^\nu + \chi_W^a W_E^\mu D_E \frac{\mu}{T}   \nonumber \\
&+ \chi_W^b W_E^\mu D_E \frac{1}{T} + \chi_W^c W_E^\mu \nabla^E_\nu u_E^\nu + \chi_W^d W_E^\nu \nabla^E_\nu u_E^\mu + \chi_W^e W_E^\nu \nabla_E^\mu u^E_\nu, 
\label{eq:dissipation}
\end{align}
where $\kappa_W$ is the energy conductivity, $\tau_W$ is the relaxation time of the energy dissipation, and $\chi_W$ are the second-order transport coefficients.
The mode analyses indicate that the causality and stability conditions are satisfied when $\kappa_W\geq 0$ and $\tau_W - \kappa_W/(e+P)T\geq 0$ in the long wavelength limit. 

The correspondences between the transport coefficients are obtained by identifying the entropy production $\partial_\mu s^\mu$ in the two frames. They are expressed as
\begin{align}
\kappa_V &= \kappa_W \bigg( \frac{n}{e+P} \bigg)^2, \ \
\tau_V = \tau_W - \frac{\kappa_W}{(e+P)T}, \ \
\chi_V^a = \chi_W^a - \frac{\tau_W nT}{e+P}, \nonumber \\
\chi_V^b &= \chi_W^b + \tau_W T - \frac{\kappa_W}{e+P}, \ \
\chi_V^c =  \chi_W^c + \frac{\kappa_W}{(e+P)T}, \ \
\chi_V^d = \chi_W^d + \frac{\kappa_W}{(e+P)T}, \ \
\chi_V^e = \chi_W^e . \label{rel}
\end{align}
They suggest that the causality and stability conditions in the Landau and the Eckart frames are strongly related because the conditions set on $\kappa_V$ and $\tau_V$ are equivalent to those on $\kappa_W$ and $\tau_W$. 

\section{Numerical simulations for nuclear collisions}

\begin{figure}[tb]
\includegraphics[width=.45\textwidth]{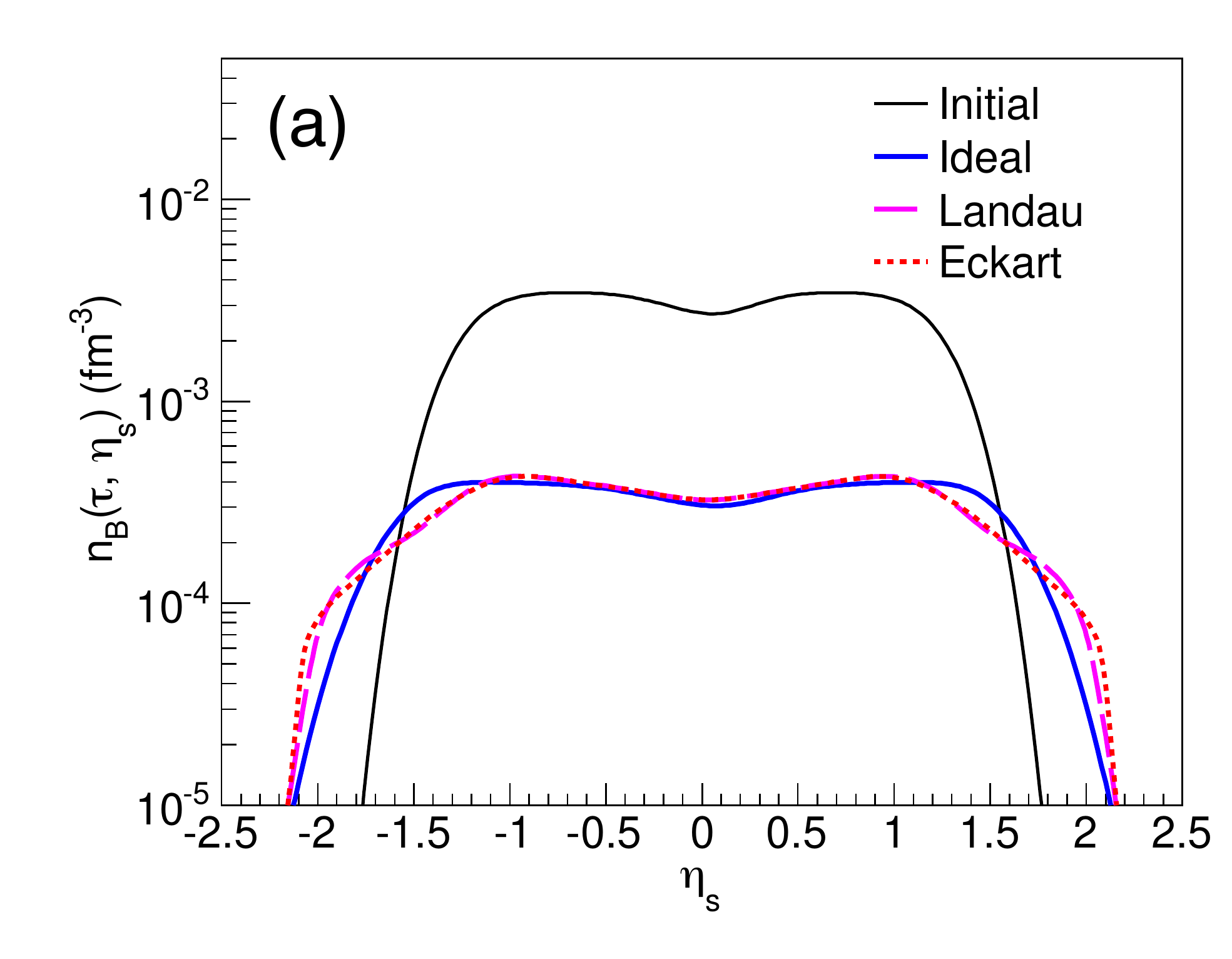}
\includegraphics[width=.45\textwidth]{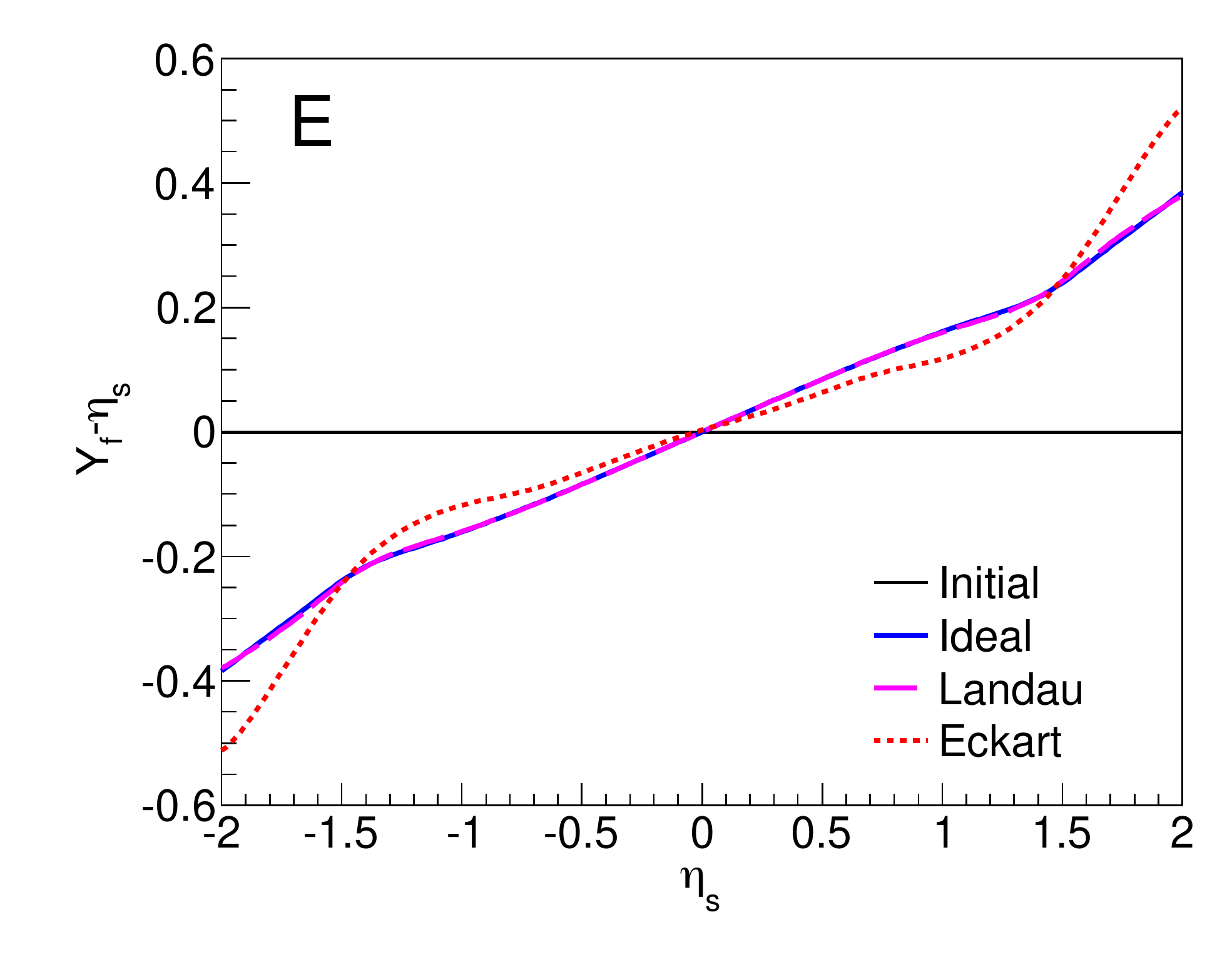} \ \\[-28pt]
\caption{(a) The net baryon density and (b) the difference between the flow and space-time rapidities at the initial time (thin solid line) compared to those during ideal (thick solid line), baryon diffusive (dashed line), and energy dissipative (dotted line) hydrodynamic evolutions at $\tau = 20$ fm/$c$.}
\label{fig:1}
\end{figure}

The hydrodynamic model of relativistic nuclear collisions is developed in the Landau and Eckart frames in a (1+1)-dimensional geometry for numerical comparison. The conserved charge in the system is the net baryon number. 
The initial conditions at $\tau_\mathrm{th}=3$ fm are parametrically constructed in accordance with the SPS data of 17.3 GeV Pb+Pb collisions. The equation of state is from \textsc{neos} B \cite{Monnai:2019hkn,Monnai:2021kgu}. $\kappa_W = 10 (e+P)$, $\tau_W = 2 \kappa_W / (e+P)T$, and $\chi_W^{a,b,c,d,e} = 0$ are used in the Eckart frame and converted for the use in the Landau frame with the relations (\ref{rel}).

\begin{figure}[tb]
\includegraphics[width=.45\textwidth]{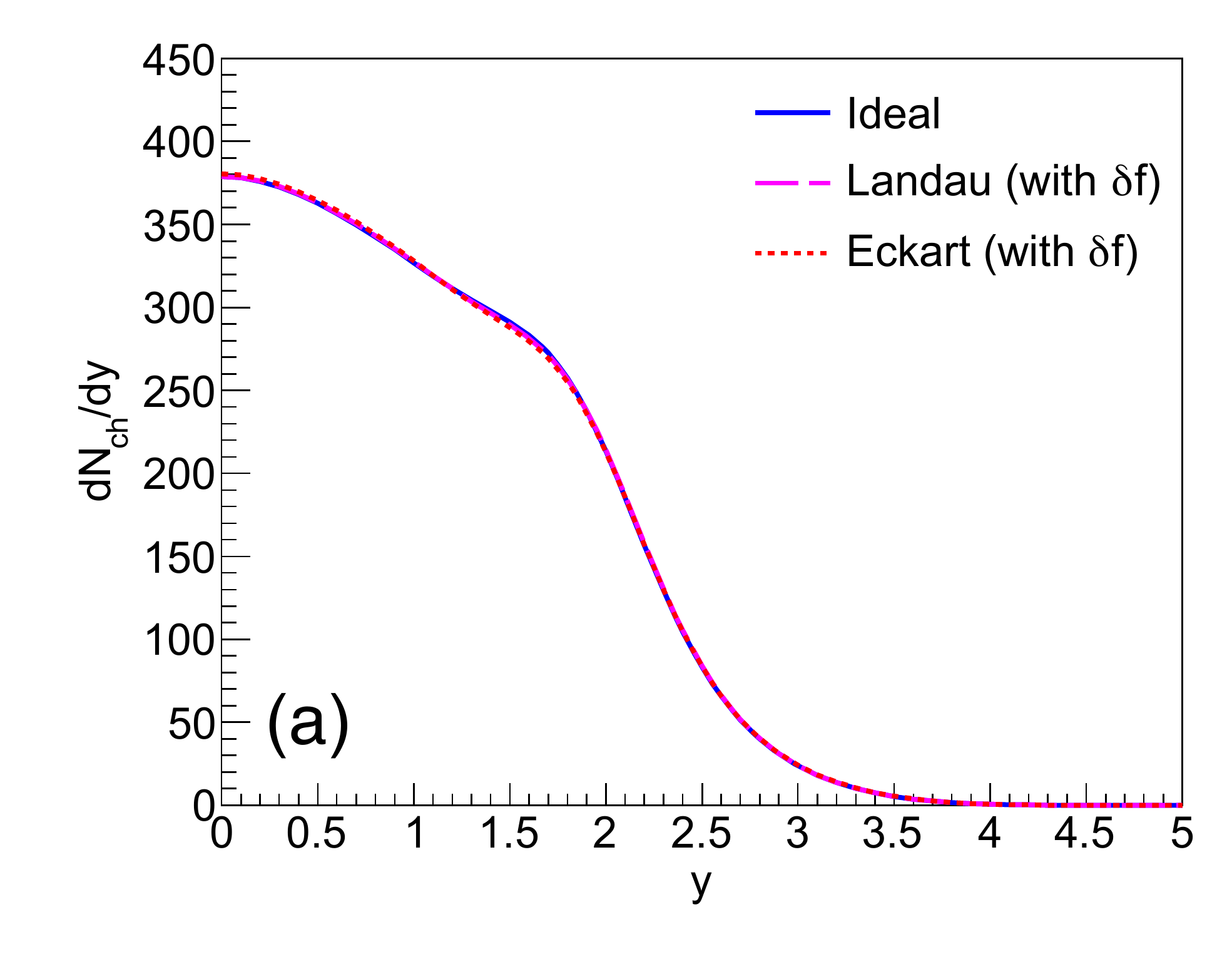}
\includegraphics[width=.45\textwidth]{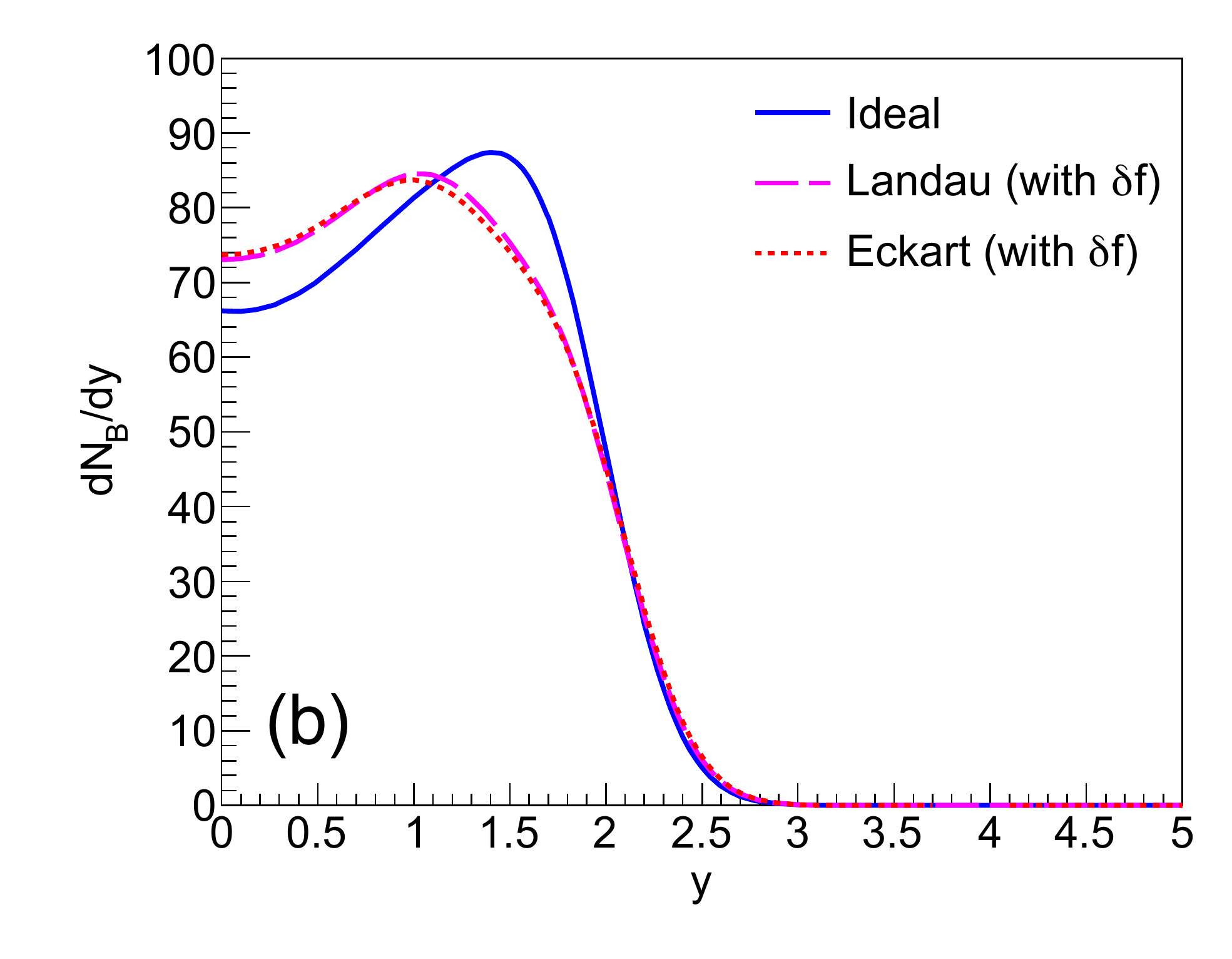} \ \\[-28pt]
\caption{The rapidity distributions of (a) charged particles and (b) net baryon number at freeze-out after ideal hydrodynamic evolution (solid line) compared to those after  baryon diffusive evolution in the Landau frame (dashed line) and energy dissipative evolution in the Eckart frame (dotted line).}
\label{fig:2}
\end{figure}

Figure \ref{fig:1} shows the space-time rapidity distributions of the net baryon density and the deviation of the flow from boost-invariance at $\tau = 20$ fm. The latter quantity is defined using the flow rapidity $Y_f$ as $Y_f-\eta_s$. The effects of the baryon diffusion and the energy dissipation on the net baryon density distribution are similar, exhibiting little frame dependence. On the other hand, the flow is shown to be sensitive to the frame choice. Figure \ref{fig:2} illustrates the rapidity distributions of charged particles and net baryon number calculated using the Cooper-Frye formula with off-equilibrium corrections to the phase-space distributions \cite{Monnai:2010qp}. It indicates that the latter is visibly affected by diffusion and dissipation processes, though the frame dependence on the observable would be small. 

\section{Discussion and summary}

The stability and causality conditions are investigated for the full second-order relativistic hydrodynamics in the Landau and Eckart frames. Numerical analyses for heavy-ion collisions imply that a frame choice has visible effects on the hydrodynamic flow but not on the particle distributions.
It is note-worthy that the second-order accuracy is required for the discussion of stability and causality even in the first-order theories because the first-order terms may have second-order differences that could be interpreted as a relaxation term when the identity
$(e+P)Du^\mu = \nabla^\mu P
- W^\mu \nabla_\nu u^\nu - W^\nu \nabla_\nu u^\mu - \Delta^{\mu \nu} DW_\nu 
$ is used \cite{Monnai:2019jkc}.
Future prospects include analyses of boosted systems, interplay with shear and bulk viscosities and (3+1)D hydrodynamic evolution.

\end{document}